\documentclass[11pt,a4paper]{article}
\usepackage{amsmath,amsthm,amssymb,epsfig,latexsym}
\newcommand{\be}[1]{\begin{equation}\label{#1}}
\newcommand{\ba}[1]{\begin{eqnarray}\label{#1}}
\newcommand{\ee}{\end{equation}}
\newcommand{\ea}{\end{eqnarray}}
\newcommand{\non}{\nonumber\\\rule{0pt}{30pt}}

\newcommand{\dis}{\displaystyle}
\newcommand{\eq}[1]{(\ref{#1})}
\newcommand{\tr}{\mathop{\rm tr}}
\newcommand{\Res}{\mathop{\rm Res}}

\newcommand{\bra}[1]{\langle{#1}|}
\newcommand{\ket}[1]{|{#1}\rangle}
\newtheorem{thm}{Theorem}[section]
\newtheorem{prop}{Proposition}[section]
\newtheorem{lemma}{Lemma}[section]

\newtheorem{Def}{Definition}[section]
\makeatletter
\@addtoreset{equation}{section}
\makeatother

\begin{document}
\begin{flushright}
LPENSL-TH-02/04\\
\end{flushright}
\par \vskip .1in \noindent

\vspace{24pt}

\begin{center}
\begin{LARGE}
{\bf Dynamical correlation functions of the $XXZ$ spin-$1/2$ chain}
\end{LARGE}

\vspace{50pt}

\begin{large}

{\bf N.~Kitanine}\footnote[1]{LPTM, UMR 8089 du CNRS, Universit\'e de Cergy-Pontoise,
France, kitanine@ptm.u-cergy.fr\par
\hspace{1.5mm}
On leave of absence from Steklov Institute at
St. Petersburg, Russia},~~
{\bf J.~M.~Maillet}\footnote[2]{ Laboratoire de Physique, UMR 5672 du CNRS,
ENS Lyon,  France,
 maillet@ens-lyon.fr},~~
{\bf N.~A.~Slavnov}\footnote[3]{ Steklov Mathematical Institute,
Moscow, Russia, nslavnov@mi.ras.ru},~~
{\bf V.~Terras}\footnote[4]{LPMT, UMR 5825 du CNRS,
Montpellier, France, terras@lpm.univ-montp2.fr} \par

\end{large}

\vspace{80pt}

\centerline{\bf Abstract} \vspace{1cm}
\parbox{12cm}{\small We derive a master equation for the dynamical spin-spin
correlation functions of the $XXZ$ spin-$\frac12$ Heisenberg
finite chain in an external magnetic field. In the thermodynamic
limit, we obtain their multiple integral representation.}
\end{center}

\newpage

\section{Introduction}

The aim of this article is to describe a new method to compute
exact representations of time-dependent correlation functions in
quantum integrable lattice models. For that purpose, we consider
the example of the $XXZ$ spin-$\frac12$ Heisenberg chain in an
external magnetic field. For simplicity, we will here mainly focus on the
dynamical correlation function of the third component of spin. It
will be clear however that the method is general and can be
applied to other cases as well.

This work is a continuation of \cite{KitMST04a}, where
we have derived a master equation for the (time-independent)
correlation functions of the $XXZ$  chain.
In the end of that article, we have
announced a multiple integral representation for the dynamical
$\sigma^z$ correlation function. We give here a proof of
that result.

The Hamiltonian of the periodical $XXZ$ spin-$\frac12$
Heisenberg chain in a magnetic field \cite{Hei28} is given by
\be{IHamXXZ}
H=H^{(0)}-hS_z,
\ee
where
\be{H-0-Sz}
\begin{array}{l}
{\dis H^{(0)}=\sum_{m=1}^{M}\left(
\sigma^x_{m}\sigma^x_{m+1}+\sigma^y_{m}\sigma^y_{m+1}
+\Delta(\sigma^z_{m}\sigma^z_{m+1}-1)\right),}\non
{\dis S_z=\frac{1}{2}\sum_{m=1}^{M}\sigma^z_{m}, \qquad
[H^{(0)},S_z]=0.}
\end{array}
\ee
Here $\Delta$ is the anisotropy parameter, $h$ denotes the
external classical magnetic field, and $\sigma^{x,y,z}_{m}$ are the
spin operators (in the spin-$\frac12$ representation) associated
with each site of the chain. The length of the chain $M$ is chosen
to be  even. The simultaneous reversal of all spins is equivalent
to a change of  sign of the magnetic field, therefore it is enough
to consider the case $h\ge 0$. The quantum space of states
${\cal H}$ is ${\cal H}={\otimes}_{m=1}^M {\cal H}_m$, where
${\cal H}_m\sim \mathbb{C}^2$ is called local quantum space. The
operators $\sigma^{x,y,z}_{m}$ act as the corresponding Pauli
matrices in the space ${\cal H}_m$ and as the identity operator
elsewhere.

The time-dependent local spin operators are defined as
\be{Def-time-dep}
\sigma^{x,y,z}_{m}(t)=e^{iHt}\sigma^{x,y,z}_{m}e^{-iHt}.
\ee
Since $[\sigma_m^z,S_z]=0$, we have for the local operator of the
third components of spin
\be{time-sig-z}
\sigma^{z}_{m}(t)=e^{iH^{(0)}t}\sigma^{z}_{m}e^{-iH^{(0)}t}.
\ee
Hence, the dynamical two-point $\sigma^{z}$ correlation function at
zero temperature is given as the following mean-value:
\be{Defzzt} g_{z z}(m,t)=\bra{\psi_g}\sigma^z_{1} e^{iH^{(0)}t}
\sigma^z_{m+1} e^{-iH^{(0)}t}\ket{\psi_g},
\ee
where $\ket{\psi_g}$ denotes the ground state of the Hamiltonian
(\ref{IHamXXZ}).

The method to compute eigenstates and energy levels (Bethe
ansatz) of the Hamiltonian \eq{IHamXXZ} was proposed by Bethe in
1931 in \cite{Bet31} and developed later in
\cite{Orb58,Wal59,YanY66a}. The algebraic version of the Bethe
ansatz was created in the framework of the Quantum Inverse
Scattering Method by L.D. Faddeev and his school
\cite{FadST79,TakF79,BogIK93L}. Different ways to study the
time-independent correlation functions  of this model were
proposed in the series of works (see e.g.
\cite{JimMMN92,JimM96,JimML95,KitMT99,
KitMT00,KitMST02a,KitMST02b,KitMST02c,KitMST02d,BosKS02,GohKS04,BosJMST04}).
As for the dynamical correlation functions, up to now, the only
known exact results concern the case of free fermions $\Delta=0$
\cite{LieSM61,Nie67,Mcc68,MccBA71,Per77,SatMJ78,MccPS83,ColIKT93}.

Our aim is to obtain a multiple integral representation of the
expectation value \eq{Defzzt} for arbitrary $\Delta>-1$.
For this purpose we use two different approaches leading (hopefully) to
the same answer. Both of them are based on the algebraic
Bethe ansatz.

The first method \cite{KitMT99,KitMT00,KitMST02b}  consists in
acting with the operators $\sigma_1^z$ and $\sigma_{m+1}^z(t)$ on
the ground state $\bra{\psi_g}$ to produce a new state, say
$\bra{\psi(z,m,t)}$. Then one can compute the resulting scalar
product $\langle\psi(z,m,t)|\psi_g\rangle$. In the
time-independent case this can be done by the algebraic Bethe
ansatz and the explicit solution of the quantum inverse scattering
problem for the $\sigma_z$  operators in site $1$ and $m+1$ (see
\cite{KitMT99,MaiT99}). However, for the calculation of the
dynamical correlation functions, we also need to compute  the
action of $e^{iH^{(0)}t}$ on arbitrary states of the chain. To
achieve this we use the fact that the Hamiltonian $H^{(0)}$ can be
constructed as the logarithmic derivative of the quantum transfer
matrix ${\cal T}(\lambda)$ :
\be{ABATI} H^{(0)}=2\sinh\eta\left.\frac{d{\cal
T}(\lambda)}{d\lambda} {\cal
T}^{-1}(\lambda)\right|_{\lambda=\frac{\eta}{2}}-2M\cosh\eta,\qquad
\mbox{with}\qquad \cosh\eta=\Delta. \ee
Hence, from the Trotter type formula \cite{Tro59,Suz66}, we can obtain the
evolution operator $e^{iH^{(0)}t}$ as the following limit:
\be{lim-U}
   e^{\pm it(H^{(0)} +2M\cosh\eta)}=\lim_{L\to\infty}
         \left({\cal T}\Bigl(\frac{\eta}{2}+\varepsilon\Bigr)\cdot
         {\cal T}^{-1}\Bigl(\frac{\eta}{2}\Bigr)\right)^{\pm L},\qquad
   \varepsilon=\frac{1}{L}2it\sinh\eta.
\ee
It is moreover possible to express ${\cal T}^{-1}(\eta/2)$
in terms of ${\cal T} (-\eta/2)$. Therefore the action of
$e^{iH^{(0)}t}$ reduces to the one of some product of transfer
matrices. This enables us to apply
directly the results of \cite{KitMST02a} and
\cite{KitMST04a}, where we computed the action of any product of
transfer matrices on arbitrary states in a compact form. In the
present case it leads  to a master equation for the dynamical
spin-spin correlation function $g_{z z}(m,t)$, or more precisely
for its generating function (see Theorem \ref{M-thm}).

The second method to handle $g_{z z}(m,t)$ is to insert between
the two $\sigma^z$ operators a sum over a complete set of
eigenstates of the Hamiltonian, leading to
\be{Defzztff} g_{z z}(m,t)=\sum_i \bra{\psi_g}\sigma^z_{1}\ket{i}
\bra{i} \sigma^z_{m+1}\ket{\psi_g} e^{i(E_i - E_0)t}, \ee
where $H^{(0)}\ket{i}=E_i \ket{i}$ and $H^{(0)}\ket{\psi_g}=E_0
\ket{\psi_g}$.
 In \cite{KitMST04a} we have
shown that for the time-independent case the master equation
method allows one to define such a sum in a precise manner.
It uses in fact a twisted version of the transfer matrix ${\cal
T}(\lambda)$ and the knowledge of determinant representations for
the form factors (e.g., corresponding to the matrix elements
$\bra{\psi_g}\sigma^z_{1}\ket{i} $). This method can easily be
generalized to the time-dependent case. So, using the technique
developed in \cite{KitMST04a}, we are able to re-sum
completely the form factor expansion for the dynamical correlation
function and to show that it leads indeed to the time-dependent
master equation obtained by the first method described above.

This time-dependent master equation gives  a representation of the
dynamical $\sigma^z$ correlation function of a finite $XXZ$ chain.
In the thermodynamic limit $M\to\infty$, the model exhibits three
different regimes depending on the value of the anisotropy
parameter $\Delta$ and on the magnetic field $h$. In this limit,
following \cite{KitMT99,KitMT00,KitMST02a,KitMST04a}, we obtain a
multiple integral representation for the dynamical $\sigma^z$
correlation function in the massive ($\Delta > 1$) and massless
($-1 < \Delta \le 1$) regimes. It is given as a sum of multiple
integrals (see \eq{answer}) involving the time dependence through
the factors $e^{iEt}$, where $E$ is the bare one-particle energy.

This paper is organized as follows. In Section 2 we introduce some
useful definitions and notations
and recall  briefly the properties of the twisted transfer matrix.
The time-dependent master equation \eq{Master-2} for the generating function of
the dynamical $\sigma^z$ correlation function is given in  Section 3.
In the following two sections we present two independent proofs of \eq{Master-2}.
The first one uses the Trotter--Suzuki procedure for the explicit calculation
of the action of $e^{iH^{(0)}t}$ on arbitrary state of the chain (Section 4).
In Section 5 we derive the time-dependent master equation via the
form factor type expansion. In Section 6 we obtain a multiple
integral representation for the dynamical $\sigma^z$ correlation function
in the thermodynamic limit.

Some technical details are gathered in two appendices. In
Appendix \ref{AS-TBE} we prove several lemmas on the properties
of the solutions of the twisted Bethe equations and of the
corresponding eigenstates of the twisted transfer matrix. These
lemmas are used in Sections 4 and 5.  In Appendix \ref{free-f}
we compare the representation \eq{answer} for the dynamical
$\sigma^z$ correlation function at $\Delta = 0$ with the one
obtained in \cite{ColIKT93} in the case of free fermions.
%
%Somewhere in the text of the paper we give a telephone number,
%where the first five attentive readers may call and ask for special
%prizes.
%

          %%%%%%%%%%%%%    %%       %%     %%%%%%%%%%
          %%               %%%      %%     %%        %%
          %%               %%%%     %%     %%         %%
          %%%%%%%%         %%  %%   %%     %%         %%
          %%               %%   %%  %%     %%         %%
          %%               %%     %%%%     %%         %%
          %%               %%      %%%     %%        %%
          %%%%%%%%%%%%%    %%       %%     %%%%%%%%%%

\section{Algebraic Bethe ansatz and twisted transfer matrix}\label{B-f}

To derive the master equation for the dynamical correlation functions we
use the twisted transfer matrix in the framework of algebraic Bethe ansatz.
The definition of this operator and related
objects together with a brief sketch of their properties is given below.
We refer the reader for more details to \cite{KitMST04a}.

\subsection{General framework}

The central object of algebraic Bethe ansatz is the
quantum monodromy matrix.
In the case of the $XXZ$ chain \eq{IHamXXZ}, this is  a $2\times 2$ matrix
\be{ABAT}
T(\lambda)=\left(
\begin{array}{cc}
A(\lambda)&B(\lambda)\\
C(\lambda)&D(\lambda)
\end{array}\right)
\ee
with operator-valued entries $A,B,C$ and $D$ which depend on
a complex parameter $\lambda$
and act in the quantum space of states ${\cal H}$ of the chain.
These operators satisfy a set of quadratic commutation relations given by
the $R$-matrix of the model.

In the framework of algebraic Bethe ansatz,
a sector ${\cal H}^{(M/2-N)}$ of the space of states $\cal H$
with a fixed number $N$ of spin down,
$N=0,1,\dots, M$, is spanned by vectors of the form
\be{ABAES}
|\psi\rangle=\prod_{j=1}^{N}B(\lambda_j)|0\rangle,
\ee
where $|0\rangle$ is the state with all spins up and
$\lambda_1,\dots,\lambda_N$ are arbitrary complex numbers.
The dual space is constructed in terms of
\be{ABADES}
\langle\psi|=\langle 0|\prod_{j=1}^{N}C(\lambda_j),
\ee
where  $\langle 0|=|0\rangle^{+}$.

The Hamiltonian $H^{(0)}$ of the homogeneous $XXZ$ chain
is given by
the `trace identity' \eq{ABATI} in terms of the transfer matrix
${\cal T}(\lambda)=A(\lambda)+D(\lambda)$.
All these transfer matrices commute, $[{\cal T}(\lambda),{\cal T}(\mu)]=0$, and
their common eigenstates
(which coincide with those of the Hamiltonian)
are usually constructed in the form \eq{ABAES} (or \eq{ABADES}) for
$\lambda_1,\dots,\lambda_N$ satisfying a system of $N$ algebraic equations
first obtained by Bethe in \cite{Bet31}.

The local spin operators can also be expressed in terms of the entries
of the monodromy matrix,
by solving the quantum inverse scattering problem \cite{KitMT99,MaiT99}:
\be{FCtab}
\sigma^\alpha_j={\cal T}^{j-1}(\eta/2)\cdot
\tr\bigl(T(\eta/2)\sigma^\alpha\bigr)
\cdot{\cal T}^{-j}(\eta/2).
\ee
Here $\sigma^\alpha$ in the r.h.s. acts in the auxiliary space of
$T(\lambda)$, while $\sigma^\alpha_j$ in the l.h.s. acts in the
local quantum space ${\cal H}_j$.

\subsection{Twisted transfer matrix}

It is actually convenient to introduce a slightly more general object:

\begin{Def}\label{Tw-T-M}
The operator
\be{Twis-T-M}
{\cal T}_\kappa(\mu)=A(\mu)+\kappa D(\mu),
\ee
where $\kappa$ is a complex parameter, is called the twisted transfer matrix.
The particular case of ${\cal T}_\kappa(\mu)$ at $\kappa=1$ corresponds to
the usual transfer matrix ${\cal T}(\mu)$.
\end{Def}

Let us consider the action of the operator ${\cal T}_\kappa(\mu)$ in the
subspace ${\cal H}^{(M/2-N)}$ with fixed (but arbitrary) number
of spins down $N$. The eigenstates of ${\cal T}_\kappa(\mu)$
and their dual states in this subspace  are denoted
as $|\psi_\kappa(\{\lambda\})\rangle$ (respectively
$\langle\psi_\kappa(\{\lambda\})|$). They can be written in the form
\eq{ABAES} (respectively \eq{ABADES}), where the
parameters $\lambda_1,\dots,\lambda_N$ satisfy the
system of twisted Bethe equations
\be{TTMBE_Y}
{\cal Y}_\kappa(\lambda_j|\{\lambda\})=0, \qquad j=1,\dots,N,
\ee
where
\be{TTM_Y-def}
{\cal Y}_\kappa(\mu|\{\lambda\}) =
a(\mu)\prod_{k=1}^{N}\sinh(\lambda_k-\mu+\eta)
+ \kappa\,d(\mu)
\prod_{k=1}^{N}\sinh(\lambda_k-\mu-\eta),
\ee
and
\be{ad}
a(\mu)=\sinh^M(\mu+\eta/2),\qquad
d(\mu)=\sinh^M(\mu-\eta/2).
\ee
The corresponding eigenvalue of
${\cal T}_\kappa(\mu)$ on $|\psi_\kappa(\{\lambda\})\rangle$ (or
on a dual eigenstate) is
\be{ABAEV}
\tau_\kappa(\mu|\{\lambda\})=
a(\mu)\prod_{k=1}^{N}\frac{\sinh(\lambda_k-\mu+\eta)}{\sinh(\lambda_k-\mu)}
+ \kappa\,d(\mu)
\prod_{k=1}^{N}\frac{\sinh(\mu-\lambda_k+\eta)}{\sinh(\mu-\lambda_k)}.
\ee
Note that
\be{TTM_Y-funct}
{\cal Y}_\kappa(\mu|\{\lambda\}) =
\prod_{k=1}^{N}\sinh(\lambda_k-\mu)\cdot
\tau_\kappa(\mu|\{\lambda\}).
\ee

\begin{Def}
A solution $\{\lambda\}$ of the system \eq{TTMBE_Y} is called {\sl admissible} if
\be{admiss}
d(\lambda_j)\prod_{k=1\atop{k\ne j}}^N\sinh(\lambda_j-\lambda_k+\eta)\ne0,\qquad
j=1,\dots,N,
\ee
and {\sl unadmissible} otherwise. A solution is called {\sl off-diagonal} if
the parameters $\lambda_1,\dots,\lambda_N$ are pairwise distinct,
and {\sl diagonal} otherwise.
\end{Def}

It is proven in Appendix A (see Theorem \ref{basis}) that the
eigenstates corresponding to the admissible off-diagonal
solutions of the system \eq{TTMBE_Y} form a basis in ${\cal H}^{(M/2-N)}$,
at least
if $\kappa$ is in a punctured vicinity of the origin (i.e. $0<|\kappa|<\kappa_0$).

{\sl Notations}. Recall that in the particular case $\kappa=1$
the corresponding twisted transfer matrix is simply denoted by ${\cal T}$
in which the subscript $\kappa$ has been omitted.
We follow the same agreement for all related
objects. Namely, $(\psi,\tau,{\cal Y})=
(\psi_\kappa,\tau_\kappa,{\cal Y}_\kappa)|_{\kappa=1}.$

At $\kappa=1$, it follows from the trace identity \eq{ABATI} that
the eigenstates of the transfer matrix ${\cal T}$ coincide
with the ones of the Hamiltonian \eq{IHamXXZ}. The corresponding eigenvalues
can be obtained from \eq{ABATI}, \eq{ABAEV}:
\be{eigen-val-H}
H^{(0)}|\psi(\{\lambda\})\rangle=\biggl(\sum_{j=1}^N E(\lambda_j)\biggr)\cdot
|\psi(\{\lambda\})\rangle,
\ee
where $E(\lambda)$ is called the bare one-particle energy and is equal to
\be{E}
E(\lambda)=\frac{2\sinh^2\eta}
{\sinh(\lambda+\frac{\eta}{2})\sinh(\lambda-\frac{\eta}{2})}.
\ee
One can similarly define the bare
one-particle momentum. It is given by
\be{P}
p(\lambda)=i\log\left(\frac{\sinh(\lambda-\frac{\eta}{2})}
{\sinh(\lambda+\frac{\eta}{2})}\right).
\ee

{\sl Remark}. The equation \eq{eigen-val-H} holds if the parameters
$\{\lambda\}$ of the eigenstate $|\psi(\{\lambda\})\rangle$ correspond to
an admissible off-diagonal solution of the system \eq{TTMBE_Y} at $\kappa=1$:
\be{BE_Y}
{\cal Y}(\lambda_j|\{\lambda\})\equiv
a(\lambda_j)\prod_{k=1}^{N}\sinh(\lambda_k-\lambda_j+\eta)
+ d(\lambda_j)\prod_{k=1}^{N}\sinh(\lambda_k-\lambda_j-\eta)=0, \qquad j=1,\dots,N.
\ee
Unlike for generic $\kappa$,  the eigenstates basis  of
${\cal T}$ (i.e. the eigenstates basis of the Hamiltonian) includes
also states corresponding to unadmissible solutions
of the system \eq{BE_Y}. We do not
consider the expectation values of the operators with respect to such states.

\subsection{Scalar products}

We recall here the expressions for the scalar product of an
eigenstate  of the twisted transfer matrix with any arbitrary
state of the form \eq{ABAES} or \eq{ABADES}.

Let us first introduce some notations.
We define, for arbitrary positive integers $n,n'$ ($n\le n'$) and
sets of variables $\lambda_1,\dots,\lambda_n$, $\mu_1,%
\dots,\mu_n$ and $\nu_1,\dots,\nu_{n'}$
such that $\{\lambda\} \subset \{\nu\}$,  the $n\times n$ matrix
$\Omega_\kappa(\{\lambda\},\{\mu\}|\{\nu\})$  as
\begin{multline} \label{matH}
  (\Omega_\kappa)_{jk}(\{\lambda\},\{\mu\}|\{\nu\})=
  a(\mu_k)\,t(\lambda_j,\mu_k)\,\prod_{a=1}^{n'} \sinh(\nu_a-\mu_k+\eta)\\
   -\kappa\, d(\mu_k)\,t(\mu_k,\lambda_j)\,\prod_{a=1}^{n'} \sinh(\nu_a-\mu_k-\eta),
\end{multline}
with
\be{def-t}
t(\lambda,\mu)=\frac{\sinh\eta}{\sinh(\lambda-\mu)\sinh(\lambda-\mu+\eta)}.
\ee
We also define, for arbitrary sets of variables
$\lambda_1,\dots,\lambda_n$ and
$\mu_1,\dots,\mu_n$,  the Cauchy determinant  ${\cal X}_n(\{\mu\},\{\lambda\})$
as
\be{TTM_Cauchy}
{\cal X}_n(\{\mu\},\{\lambda\})\equiv\det_n\left(
\frac1{\sinh(\mu_k-\lambda_j)}\right)=
\frac{\prod\limits_{a>b}^n\sinh(\lambda_a-\lambda_b)
\sinh(\mu_b-\mu_a)}{\prod\limits_{a,b=1}^n\sinh(\mu_b-\lambda_a)}.
\ee

\begin{prop}\cite{Sla89,KitMT99}
Let $\lambda_1,\dots,\lambda_N$ satisfy the system \eq{TTMBE_Y},
$\mu_1,\dots,\mu_N$ be generic complex numbers. Then
\begin{align}
\langle 0|\prod_{j=1}^{N}C(\mu_j)|\psi_\kappa(\{\lambda\})\rangle
&=
\langle\psi_\kappa(\{\lambda\})|\prod_{j=1}^{N}B(\mu_j)|0\rangle
\hspace{5.5cm}\nonumber
\\
&=  \prod_{a=1}^{N} d(\lambda_a)\cdot
    {\cal X}_N^{-1}(\{\mu\},\{\lambda\})\cdot
   \det_N \left(\frac\partial{\partial\lambda_j}
   \tau_\kappa(\mu_k|\{\lambda\})\right)\label{FCdet}
\\
&=  \frac{\prod_{a=1}^{N} d(\lambda_a)}
{\prod\limits_{a>b}^N\sinh(\lambda_a-\lambda_b)\sinh(\mu_b-\mu_a)}
\cdot \det_N
\Omega_\kappa(\{\lambda\},\{\mu\}|\{\lambda\})\label{scal-prod}.
\end{align}
\end{prop}

The eigenstate $|\psi_\kappa(\{\lambda\})\rangle$ is orthogonal to the
dual eigenstate $\langle\psi_\kappa(\{\mu\})|$, if the sets $\{\lambda\}$
and $\{\mu\}$ are different: $\{\lambda\}\ne\{\mu\}$ (see
Appendix \ref{AS-TBE}). Otherwise
\begin{align}\label{Y-Jac}
\langle \psi_\kappa(\{\lambda\})|\psi_\kappa(\{\lambda\})\rangle
&= \frac{\prod_{a=1}^{N} d(\lambda_a)}
{\prod\limits_{a,b=1\atop{a\ne b}}^N\sinh(\lambda_a-\lambda_b)}
\cdot \det_N \Omega_\kappa(\{\lambda\},\{\lambda\}|\{\lambda\})
\\
&=(-1)^N\frac{\prod_{a=1}^{N} d(\lambda_a)}
{\prod\limits_{a,b=1\atop{a\ne b}}^N\sinh(\lambda_a-\lambda_b)}
\cdot \det_N \left(\frac\partial{\partial\lambda_k}
{\cal Y}_\kappa(\lambda_j|\{\lambda\})\right).
\label{norm}
\end{align}

The equations \eq{FCdet}--\eq{norm} are valid for any arbitrary
complex parameter $\kappa$, in particular at $\kappa=1$. In this
case we denote $\Omega=\left.\Omega_\kappa\right|_{\kappa=1}$.

          %%%%%%%%%%%%%    %%       %%     %%%%%%%%%%
          %%               %%%      %%     %%        %%
          %%               %%%%     %%     %%         %%
          %%%%%%%%         %%  %%   %%     %%         %%
          %%               %%   %%  %%     %%         %%
          %%               %%     %%%%     %%         %%
          %%               %%      %%%     %%        %%
          %%%%%%%%%%%%%    %%       %%     %%%%%%%%%%

\section{Master equation for dynamical correlation
functions}\label{DNK}

Our main goal is to obtain an explicit representation for
the time dependent correlation function of the third components
of spin,
\be{exp-val-sig}
\langle \sigma_{1}^z(0)\sigma_{m+1}^z(t)\rangle=
\frac{\langle\psi(\{\lambda\})|\sigma_{1}^z(0)\sigma_{m+1}^z(t)
|\psi(\{\lambda\})\rangle}
{\langle\psi(\{\lambda\})|\psi(\{\lambda\})\rangle}.
\ee
In this expression, $|\psi(\{\lambda\})\rangle$ denotes the
ground state of the Hamiltonian \eq{IHamXXZ} in the subspace
${\cal H}^{(M/2-N)}$. Note however that all the results we
present here for the finite chain remain valid for $\{\lambda\}$
being more generally any admissible off-diagonal solution of the
system \eq{BE_Y}.

Like in \cite{KitMST04a}, it is convenient to derive this correlation function
from a special generating function.
Indeed, let us consider the following time-dependent operator:
\be{Qkm}
Q^\kappa_{l+1,m}(t)
   ={\cal T}^{l}\Bigl(\frac{\eta}{2}\Bigr)\cdot
    {\cal T}^{m-l}_\kappa\Bigl(\frac{\eta}{2}\Bigr)\cdot
    e^{ itH^{(0)}_\kappa}\cdot
    {\cal T}^{-m}\Bigl(\frac{\eta}{2}\Bigr)\cdot e^{- itH^{(0)}},
\ee
with
$H^{(0)}_\kappa$ defined similarly as in \eq{ABATI}:
\be{ABATI-kap}
   H_\kappa^{(0)}=
     2\sinh\eta\left.\frac{d{\cal T_\kappa}(\lambda)}{d\lambda}
     {\cal T}_\kappa^{-1}(\lambda)\right|_{\lambda=\frac{\eta}{2}}
     -2M\cosh\eta.
\ee
Using the expression \eq{FCtab} for the reconstruction of the local spin
operators in terms of the entries of the monodromy matrix and the fact that
the twisted transfer matrix ${\cal T}_\kappa$ commutes with $H^{(0)}_\kappa$,
it is easy to see
that
\begin{align}
  \frac{1-\sigma_{l+1}^z(0)}{2}\cdot\frac{1-\sigma_{m+1}^z(t)}{2}
     &= {\cal T}^{l}\Bigl(\frac{\eta}{2}\Bigr)\cdot
        D\Bigl(\frac{\eta}{2}\Bigr)\cdot
        {\cal T}^{m-l-1}\Bigl(\frac{\eta}{2}\Bigr)\cdot
        e^{ itH^{(0)}}\nonumber\\
     &\hspace{4.3cm}\times
        D\Bigl(\frac{\eta}{2}\Bigr)\cdot
        {\cal T}^{-m-1}\Bigl(\frac{\eta}{2}\Bigr)\cdot
        e^{- itH^{(0)}},\\
     &= \frac{1}{2} \frac{\partial^2}{\partial\kappa^2}
        ( Q^\kappa_{l+1,m+1} - Q^\kappa_{l+1,m} -
          Q^\kappa_{l+2,m+1} + Q^\kappa_{l+2,m})(t) |_{\kappa=1}.
     \label{2lat-der}
\end{align}
Due to the translational invariance of the correlation functions,
we can set $l=0$ and simply consider the following expectation value:
\be{def-Q}
{\cal Q}_\kappa(m,t)=
\frac{\langle\psi(\{\lambda\})|Q^\kappa_{1,m}(t)
|\psi(\{\lambda\})\rangle}
{\langle\psi(\{\lambda\})|\psi(\{\lambda\})\rangle}.
\ee
In terms of this generating function,
the time-dependent correlation function \eq{exp-val-sig}
is thus given  as
\be{cor-funct-ss}
\langle\sigma_{1}^z(0)\sigma_{m+1}^z(t)\rangle
=2\langle\sigma_{1}^z(0)\rangle-1+
2D^2_m\left.\frac{\partial^2}{\partial\kappa^2}
{\cal Q}_\kappa(m,t)\right|_{\kappa=1},
\ee
where $D^2_m$ denotes the second lattice derivative defined as in
\eq{2lat-der}.

Like in the time-independent case, it is possible to derive a master equation
for the generating function \eq{def-Q} in the finite chain.
Indeed, we have the following result:

\begin{thm}\label{M-thm}
Let $\{\lambda_1,\ldots,\lambda_N\}$ be an admissible off-diagonal solution
of the system \eq{BE_Y}. Then there exists $\kappa_0 > 0$ such that,
for $0<|\kappa|< \kappa_0 $,
the generating function ${\cal Q}_\kappa(m,t)$ \eq{def-Q}
in the finite $XXZ$ chain \eq{IHamXXZ}
is given by the multiple contour integral
\begin{multline}\label{Master-2}
{\cal Q}_\kappa(m,t) =  \frac{1}{N!}
 \oint\limits_{\Gamma\{\pm\frac{\eta}2\}\cup \Gamma\{\lambda\}}
 \prod_{j=1}^{N}\frac{dz_j}{2\pi i} \cdot
 \prod_{b=1}^{N}e^{it\bigl(E(z_b)-E(\lambda_b)\bigr)
+im\bigl(p(z_b)-p(\lambda_b)\bigr)}
\\
\times\prod_{a,b=1}^{N}\sinh^2(\lambda_a-z_b)\cdot
\frac{\det_N
\left(\frac{\partial\tau_\kappa(\lambda_j|\{z\})}{\partial z_k}\right)
\cdot\det_N
\left(\frac{\partial\tau(z_k|\{\lambda\})}{\partial \lambda_j}\right)}
{\prod\limits_{a=1}^{N}{\cal Y}_\kappa (z_a| \{z\} )
\cdot\det_N
\left(\frac{\partial{\cal Y}(\lambda_k|\{\lambda\})}
{\partial\lambda_j}\right)}.
\end{multline}
In this expression, $E(\lambda)$ and $p(\lambda)$ denote respectively
the bare one-particle energy and momentum  \eq{E} and \eq{P};
the integration contour is such that the only singularities of
the integrand \eq{Master-2} within
$\Gamma\{\pm\frac{\eta}2\}\cup \Gamma\{\lambda\}$ which contribute to the
integral  are the points $\{\pm\frac{\eta}2\}$ and $\{\lambda\}$.
\end{thm}

{\sl Remark 1}. The master equation \eq{Master-2}
gives the expectation value ${\cal Q}_\kappa(m,t)$
with respect to an arbitrary eigenstate $|\psi(\{\lambda\})\rangle$ of
${\cal T}$ corresponding to any admissible
off-diagonal solution $\{\lambda_1,\ldots,\lambda_N\}$  of
\eq{BE_Y}. In particular one can choose $\{\lambda\}$ such that
$|\psi(\{\lambda\})\rangle$ is the ground state of the
$XXZ$ Hamiltonian.

{\sl Remark 2}. The master equation \eq{Master-2} provides
an integral representation of
${\cal Q}_\kappa(m,t)$ which is valid
at least in a punctured vicinity of $\kappa=0$. On the other hand,
it is clear from \eq{Qkm} and \eq{def-Q} that
${\cal Q}_\kappa(m,t)$ is an analytical function of $\kappa$ for
$\kappa\in\mathbb{C}\setminus\{0,\infty\}$. Hence,
the representation \eq{Master-2}
can be analytically continued from any vicinity of the origin to the whole
complex plane except $\kappa=0,\infty$. This does not mean, however,
that one
can set $\kappa$ to be an arbitrary specific value directly in the
integrand of
\eq{Master-2}.

{\sl Remark 3}. The time and space dependencies  in the
representation \eq{Master-2} appear in a very suggestive way,
namely through the exponent of the bare energy and momentum. Note
however that they do not correspond {\sl a priori} to any
eigenstate since here $z_j$'s are integration variables;
nevertheless, the contribution of each point in the integration
domain is measured in particular by the difference between the
bare energy and momentum corresponding to the integration
variables $z_j$ and the one corresponding to the ground state
parameters $\lambda_j$.

\bigskip

As we already discussed in the Introduction, there are two
possible ways to prove Theorem \ref{M-thm}.

The first one, which follows the main strategy
of \cite{KitMST04a}, consists in acting with $Q^\kappa_{1,m}(t)$ on the state
$\langle\psi(\{\lambda\})|$, in computing the resulting scalar products and
in rewriting the sum over partitions that follows as a single multiple integral
of Cauchy type. In the next section, we explain how this approach, elaborated
in  \cite{KitMST04a}, can be
applied in the time-dependent case in order to prove  Theorem \ref{M-thm}.

The second one was already announced in the conclusion of \cite{KitMST04a}:
the master equation \eq{Master-2} can be obtained directly via a
form factor expansion. Details are presented in Section \ref{sec-ff}.

The reader who is not interested in the technical details of these derivations
can skip the next two sections and go directly to Section \ref{sec-th}
where a multiple integral representation of \eq{def-Q} in the thermodynamic
limit is deduced from \eq{Master-2}.

          %%%%%%%%%%%%%    %%       %%     %%%%%%%%%%
          %%               %%%      %%     %%        %%
          %%               %%%%     %%     %%         %%
          %%%%%%%%         %%  %%   %%     %%         %%
          %%               %%   %%  %%     %%         %%
          %%               %%     %%%%     %%         %%
          %%               %%      %%%     %%        %%
          %%%%%%%%%%%%%    %%       %%     %%%%%%%%%%

\section{Master equation via multiple action of transfer matrices}
\label{sec-tm}

In this section, we prove Theorem \ref{M-thm}
using the method developed in \cite{KitMST02a} and \cite{KitMST04a}.
In order  to apply this method, we need first to reduce the
computation of the generating function \eq{def-Q}  to the evaluation of the
expectation value of some product of twisted transfer matrices.
The idea is to reconstruct the operator $\exp(itH^{(0)}_\kappa)$ in
\eq{Qkm} as a  limit similar to \eq{lim-U}:
\be{lim-U-k}
   e^{it(H^{(0)}_\kappa+2M\cosh\eta)}=\lim_{L\to\infty} \left({\cal T}_\kappa
\Bigl(\frac{\eta}{2}+\varepsilon\Bigr)\cdot
         {\cal T}_\kappa^{-1}\Bigl(\frac{\eta}{2}\Bigr)\right)^{L},\qquad
   \varepsilon=\frac{1}{L}2it\sinh\eta,
\ee
and to use the fact that,
for the specific value $\lambda=\eta/2$, the inverse operator\footnote{%
One can easily obtain the equation \eq{T-inverse} by applying the product
${\cal T}_\kappa(\eta/2){\cal T}_\kappa(-\eta/2)$ to an arbitrary state
\eq{ABAES}.
}
${\cal T}^{-1}_\kappa(\eta/2)$ is proportional to
${\cal T}_\kappa(-\eta/2)$:
\be{T-inverse}
{\cal T}_\kappa^{-1}(\eta/2)=\Bigl(\kappa\,a(\eta/2)d(-\eta/2)\Bigr)^{-1}
{\cal T}_\kappa(-\eta/2).
\ee
This enables us to express the operator
$Q_{1,m}^\kappa(t)$
as the limit
\begin{align}
  Q_{1,m}^\kappa(t) &= \lim_{L\to\infty}
       {\cal T}_\kappa^L\Bigl(\frac{\eta}{2}+\varepsilon\Bigr)\cdot
       {\cal T}_\kappa^{m-L}\Bigl(\frac{\eta}{2}\Bigr)\cdot
       {\cal T}^{L-m}\Bigl(\frac{\eta}{2}\Bigr)\cdot
       {\cal T}^{-L}\Bigl(\frac{\eta}{2}+\varepsilon\Bigr),\\
  &= \lim_{L\to\infty}\kappa^{m-L} \,
        {\cal T}_\kappa^L\Bigl(\frac{\eta}{2}+\varepsilon\Bigr)\cdot
        {\cal T}_\kappa^{L-m}\Bigl(-\frac{\eta}{2}\Bigr)\cdot
        {\cal T}^{m-L}\Bigl(-\frac{\eta}{2}\Bigr)\cdot
        {\cal T}^{-L}\Bigl(\frac{\eta}{2}+\varepsilon\Bigr).
\end{align}
Acting with the product of (untwisted) inverse transfer matrices on the
eigenstate $|\psi(\{\lambda\})\rangle$, we obtain a representation
of the generating function \eq{def-Q}  in terms of an expectation value
of a product of twisted transfer matrices:
\be{gen-fun}
   {\cal Q}_\kappa(m,t)=\lim_{L\to\infty}\kappa^{m-L}\,
    \tau^{m-L}(-\eta/2|\{\lambda\})\,
    \tau^{-L}(\eta/2+\varepsilon|\{\lambda\})\,
    \langle{\cal T}_\kappa^L(\eta/2+\varepsilon)\,
           {\cal T}_\kappa^{L-m}(-\eta/2)\rangle.
\ee
It is convenient at this stage to introduce some arbitrary parameters
$\omega_1,\ldots,\omega_{2L-m}$ and to define
$(x_1,\ldots,x_{2L-m})=
(\omega_1+\varepsilon,\ldots,\omega_L+\varepsilon,
 \omega_{L+1}-\eta,\ldots,\omega_{2L-m}-\eta)$. It follows that
\be{gen-fun1}
   {\cal Q}_\kappa(m,t)=\lim_{L\to\infty}\lim_{\omega_a\to\frac{\eta}2}
    \kappa^{m-L}
    \prod_{a=1}^{2L-m}\tau^{-1}(x_a|\{\lambda\})\,
    \langle\prod_{a=1}^{2L-m}{\cal T}_\kappa(x_a)\rangle.
\ee

As soon as we have the representation \eq{gen-fun1}, we can use directly
the equation (4.15) of \cite{KitMST04a} in order to obtain a multiple
integral representation for the expectation
value $\langle\prod{\cal T}_\kappa(x_a)\rangle$. It leads to
\begin{multline}\label{master1}
   {\cal Q}_\kappa(m,t)=\lim_{L\to\infty}\lim_{\omega_a\to\frac{\eta}2}
% \langle Q_{1,m}^\kappa\rangle =
 \kappa^{m-L}\,\frac{1}{N!}
 \oint\limits_{\Gamma\{x\} \cup \Gamma\{\lambda\}}
 \prod_{j=1}^{N}\frac{dz_j}{2\pi i} \cdot
  \prod_{a=1}^{2L-m}\frac{\tau_\kappa (x_a| \{z\} )}
                         {\tau (x_a| \{\lambda\} )}
 \\
 \times
 \prod_{a=1}^{N} \frac{1}
                      { {\cal Y}_\kappa (z_a| \{z\} )}
         \cdot
 \det_N \Omega_\kappa ( \{z\} , \{\lambda\} | \{z\}  )
    \cdot
    \frac{ \det_N \Omega (\{\lambda\} , \{z\}  |\{\lambda\} )}
         { \det_N \Omega (\{\lambda\} , \{\lambda\} | \{\lambda\})},
\end{multline}
where the closed contour $\Gamma\{x\} \cup \Gamma\{\lambda\}$
surrounds the points $x_1,\dots,x_{2L-m}$  and $\lambda_1,\dots,\lambda_N$
and does not contain any other singularities of the
integrand\footnote{%
More precisely, $\Gamma\{x\}\cup\Gamma\{\lambda\}$ is the
boundary of a set of polydisks ${\cal D}(x_a,r)$ and
${\cal D}(\lambda_b,r)$ in $\mathbb{C}^N$. Namely,
$\Gamma\{x\}={\cup}_{a=1}^{2L-m}\bar{\cal D}(x_a,r)$, where
$\bar{\cal D}(x_a,r)=\{z\in\mathbb{C}^N: |z_k-x_a|=r,\quad
k=1,\dots,N\}$. The integration contour
$\Gamma\{\lambda\}$ is defined
in a similar manner. The radius $r$ is supposed to
be small enough.
}.
Formally the integrand in the equation \eq{master1} coincides
exactly with its
time-independent analogue in \cite{KitMST04a}. The difference, however, is
that the number of parameters $x_j$ eventually becomes infinite, and that one
has to take the homogeneous limit
$\omega_j\to\eta/2$ {\em before} the limit $L\to\infty$.
In this limit, some of the  $x_j$ go to $\eta/2+
\varepsilon$, while the remaining ones tend to $-\eta/2$. Thus, to be able
to proceed to this limit, one has
to verify that all the solutions of the system
\be{TBE-z}
{\cal Y}_\kappa (z_j|\{z\})\equiv
a(z_j)\prod_{k=1}^{N}\sinh(z_k-z_j+\eta)
+\kappa\,d(z_j)\prod_{k=1}^{N}\sinh(z_k-z_j-\eta)=0
\ee
which contribute to the integral \eq{master1}
are actually separated from the points $\eta/2+\varepsilon$, $-\eta/2$ and
the parameters $\{\lambda\}$. Note that, thanks to the
analyticity of $\langle Q_{1,m}^\kappa\rangle$ for
$\kappa\in\mathbb{C}\setminus\{0,\infty\}$,
it is sufficient to evaluate this generating  function
in the punctured vicinity of  $\kappa=0$. Observe also that
one can set $\eta/2+\varepsilon$ to be as close to
$\eta/2$ as necessary, since eventually we have to proceed to the limit
$L\to\infty$.

The analysis of the solutions of the system \eq{TBE-z} performed
in the article \cite{KitMST04a} was based on the results of
\cite{TarV96}.
However, most of those results were formulated in the case of
inhomogeneous Bethe equations
\be{TBE-z-inh}
\prod_{a=1}^M\sinh^M(z_j-\xi_a+\eta)\prod_{k=1\atop{k\ne j}}^{N}
\sinh(z_k-z_j+\eta)
-\kappa\,
\prod_{a=1}^M\sinh^M(z_j-\xi_a)\prod_{k=1\atop{k\ne j}}^{N}
\sinh(z_k-z_j-\eta)=0,
\ee
for generic\footnote{%
Or at least `well separated' (see \cite{TarV96} for definition).}
parameters $\xi_1,\dots,\xi_M$. We assume that {\sl all
solutions $z_j(\kappa)$ of the homogeneous system \eq{TBE-z} can be obtained
from the solutions $z_j(\kappa|\{\xi\})$ of the inhomogeneous
system \eq{TBE-z-inh} in the limit $\xi_a\to\eta/2$.} Then, just like in the
time-independent case, unadmissible and
diagonal solutions of the system \eq{TBE-z} do not contribute to the integral
\eq{master1}, for $\det \Omega_\kappa$ or $\det \Omega$ vanishes at
these points, and the only solutions that actually contribute are
admissible off-diagonal ones.
Due to Lemma \ref{adm-sol}, all of them are separated from
the points  $\pm\eta/2$
(and hence from $\eta/2+\varepsilon$ as well, for $\varepsilon$ small enough),
at least in a vicinity of $\kappa=0$ as far as $\kappa\ne 0$.
For $|\kappa|$ small enough, they are also obviously
separated from the parameters $\{\lambda\}$, which
correspond to an admissible
off-diagonal solution at $\kappa=1$. Thus,  we can formulate
\begin{lemma}\label{main-lem}
Let $\{\lambda\}$ be an
admissible off-diagonal solution to \eq{BE_Y}, and $\omega_j\to\eta/2$.
There exists $\kappa_0>0$ such that, for  $0<|\kappa|<\kappa_0$,
one can define a closed contour
$\Gamma\{\frac\eta2\}\cup\Gamma\{-\frac\eta2\}%
\cup\Gamma\{\lambda\}$ which satisfies the following properties:

1) it surrounds the points $\eta/2$,
$-\eta/2$ and $\{\lambda\}$, while
all admissible off-diagonal solutions of the system \eq{TBE-z}
are outside of this contour;

2) for $L$ large enough, the only poles which are
inside and
provide non-vanishing contribution to the integral \eq{master1}
are $z_j=\eta/2+\varepsilon$, $z_j=-\eta/2$ and $z_j=\lambda_k$;

3) for $L$ large enough, the only poles which are outside
(within a set of strips of width $i\pi$) and
provide non-vanishing contribution to the integrand of \eq{master1}
are the admissible off-diagonal solutions of the system \eq{TBE-z}.
\end{lemma}

Thanks to this Lemma, we can now take the limit $\omega_j\to\eta/2$
in \eq{master1}. It is easy to see that
\begin{multline}\label{hom-lim}
  \lim_{\omega_a\to\eta/2}
     \biggl(\kappa^{m-L}
     \prod_{a=1}^{2L-m}\frac{\tau_\kappa(x_a|\{z\})}
                            {\tau(x_a|\{\lambda\})} \biggr)
            \\
  =\prod_{b=1}^N\left(\frac{\sinh(z_b+\frac\eta2-\varepsilon)}
                           {\sinh(z_b-\frac\eta2-\varepsilon)}
    \cdot\frac{\sinh(\lambda_b-\frac\eta2-\varepsilon)}
              {\sinh(\lambda_b+\frac\eta2-\varepsilon)}\right)^{L}\cdot
    \left(\frac{\sinh(z_b-\frac\eta2)}{\sinh(z_b+\frac\eta2)}
     \cdot\frac{\sinh(\lambda_b+\frac\eta2)}{\sinh(\lambda_b-\frac\eta2)}
         \right)^{L-m}
            \\
\times
\left[1+\frac{\sinh^M\varepsilon}{\sinh^M(\eta+\varepsilon)}
\prod_{b=1}^N\frac{\sinh(\lambda_b-\frac{3\eta}2-\varepsilon)}
{\sinh(\lambda_b+\frac{\eta}2-\varepsilon)}\right]^{-L}\cdot
\left[1+\frac{\kappa\,\sinh^M\varepsilon}{\sinh^M(\eta+\varepsilon)}
\prod_{b=1}^N\frac{\sinh(z_b-\frac{3\eta}2-\varepsilon)}
{\sinh(z_b+\frac{\eta}2-\varepsilon)}\right]^L.
\end{multline}
As the contour defined in Lemma \ref{main-lem} is independent of $L$, one
can also safely  take the limit $L\to\infty$ in the integrand.
Using the definition of the
bare one-particle energy $E(\lambda)$ and momentum $p(\lambda)$ \eq{E}, \eq{P},
one obtains
\begin{multline}\label{L-lim}
\lim_{L\to\infty}\lim_{\omega_a\to\eta/2}
     \biggl(\kappa^{m-L}
     \prod_{a=1}^{2L-m}\frac{\tau_\kappa(x_a|\{z\})}
                            {\tau(x_a|\{\lambda\})} \biggr)
            \\
=\prod_{b=1}^N\exp\Bigl(it(E(z_b)-E(\lambda_b))+im(p(z_b)-p(\lambda_b))\Bigr),
\end{multline}
which leads to the following multiple integral representation for the
generating function ${\cal Q}_\kappa(m,t)$ in
the finite chain:
\begin{multline}\label{Master-1}
 {\cal Q}_\kappa(m,t) =  \frac{1}{N!}
 \oint\limits_{\Gamma\{\pm\frac{\eta}2\}\cup \Gamma\{\lambda\}}
 \prod_{j=1}^{N}\frac{dz_j}{2\pi i} \cdot
 \prod_{b=1}^{N}e^{it\bigl(E(z_b)-E(\lambda_b)\bigr)
+im\bigl(p(z_b)-p(\lambda_b)\bigr)}
 \\
 \times
 \prod_{a=1}^{N} \frac{1}
                      { {\cal Y}_\kappa (z_a| \{z\} )}
         \cdot
 \det_N \Omega_\kappa ( \{z\} , \{\lambda\} | \{z\}  )
    \cdot
    \frac{ \det_N \Omega (\{\lambda\} , \{z\}  |\{\lambda\} )}
         { \det_N \Omega (\{\lambda\} , \{\lambda\} | \{\lambda\})}.
\end{multline}
It remains to use \eq{FCdet}--\eq{norm} and we obtain the master equation
\eq{Master-2}.

          %%%%%%%%%%%%%    %%       %%     %%%%%%%%%%
          %%               %%%      %%     %%        %%
          %%               %%%%     %%     %%         %%
          %%%%%%%%         %%  %%   %%     %%         %%
          %%               %%   %%  %%     %%         %%
          %%               %%     %%%%     %%         %%
          %%               %%      %%%     %%        %%
          %%%%%%%%%%%%%    %%       %%     %%%%%%%%%%

\section{Master equation via form factor expansion}
\label{sec-ff}

The master equation \eq{Master-2} can be obtained via the form factor expansion
for the expectation value ${\cal Q}_\kappa(m,t)$. This way is of course
much shorter, for it is in
particular not necessary to construct the generating function as a
special limit of a product of twisted transfer matrices. However,
the summation over form factors requires the completeness of the set of
eigenstates of ${\cal T}_\kappa$ (see Theorem \ref{basis}).

Inserting in \eq{def-Q} the complete set of the eigenstates
$|\psi_\kappa(\{\mu\})\rangle$ of the twisted transfer matrices
between the operators ${\cal T}_\kappa^m(\frac\eta2)
\cdot\exp(itH^{(0)}_\kappa)$ and ${\cal T}^{-m}(\frac\eta2)
\cdot\exp(-itH^{(0)})$ (see \eq{Qkm}), one obtains
\be{sum-ff}
{\cal Q}_\kappa(m,t)=\sum_{\{\mu\}}
\prod_{b=1}^N e^{it\bigl(E(\mu_b)-E(\lambda_b)\bigr)
+im\bigl(p(\mu_b)-p(\lambda_b)\bigr)}
\frac{\langle\psi (\{\lambda\})|\psi_\kappa(\{\mu\})\rangle}
{\langle\psi_\kappa(\{\mu\})|\psi_\kappa(\{\mu\})\rangle}\cdot
\frac{\langle\psi_\kappa(\{\mu\})|\psi (\{\lambda\})\rangle}
{\langle\psi (\{\lambda\})|\psi (\{\lambda\})\rangle},
\ee
where the sum is taken with respect to all admissible off-diagonal
solutions $\{\mu_1,\dots,\mu_N\}$
(such that $-\pi/2<\Im (\mu_j)\le\pi/2$) of the system \eq{TTMBE_Y}. As usual,
we suppose $|\kappa|$ small enough, but not zero. We have also used that
\begin{align}\label{tau-mom}
    &i\sum_{b=1}^Np(\lambda_b)=\log\tau_\kappa(\eta/2|\{\lambda\})
                              -M\log\sinh\eta,
                  \\
      \label{eigen-val-H-k}
    &H^{(0)}_\kappa|\psi_\kappa(\{\lambda\})\rangle=
        \biggl(\sum_{j=1}^N E(\lambda_j)\biggr)\cdot
             |\psi_\kappa(\{\lambda\})\rangle,
\end{align}
which follows from \eq{ABAEV}, \eq{P} and \eq{ABATI-kap}. The scalar products
in \eq{sum-ff} can be written in terms of Jacobians via \eq{FCdet}--\eq{norm}:
\begin{multline}\label{sum-ff-Jac}
{\cal Q}_\kappa(m,t)=(-1)^N\sum_{\{\mu\}}
\prod_{a,b=1}^{N}\sinh^2(\lambda_a-\mu_b)
\cdot\prod_{b=1}^N e^{it\bigl(E(\mu_b)-E(\lambda_b)\bigr)
+im\bigl(p(\mu_b)-p(\lambda_b)\bigr)}
\\
\times\frac{\det_N
\left(\frac{\partial\tau_\kappa(\lambda_j|\{\mu\})}{\partial \mu_k}\right)
}{\det_N\left(\frac{\partial{\cal Y}_\kappa (\mu_k| \{\mu\})}
{\partial\mu_j}\right)}\cdot
\frac{\det_N
\left(\frac{\partial\tau(\mu_k|\{\lambda\})}{\partial \lambda_j}\right)}
{\det_N \left(\frac{\partial{\cal Y}(\lambda_k|\{\lambda\})}
{\partial\lambda_j}\right)}.
\end{multline}
This last sum can be presented as a contour integral in
$\mathbb{C}^N$,
\begin{multline}\label{Master-out}
{\cal Q}_\kappa(m,t) =  \frac{(-1)^N}{N!}
 \oint\limits_{\Gamma\{\mu\}}
 \prod_{j=1}^{N}\frac{dz_j}{2\pi i} \cdot
 \prod_{b=1}^{N}e^{it\bigl(E(z_b)-E(\lambda_b)\bigr)
+im\bigl(p(z_b)-p(\lambda_b)\bigr)}
\\
\times\prod_{a,b=1}^{N}\sinh^2(\lambda_a-z_b)\cdot
\frac{\det_N
\left(\frac{\partial\tau_\kappa(\lambda_j|\{z\})}{\partial z_k}\right)
\cdot\det_N
\left(\frac{\partial\tau(z_k|\{\lambda\})}{\partial \lambda_j}\right)}
{\prod\limits_{a=1}^{N}{\cal Y}_\kappa (z_a| \{z\} )
\cdot\det_N
\left(\frac{\partial{\cal Y}(\lambda_k|\{\lambda\})}
{\partial\lambda_j}\right)},
\end{multline}
where the contour $\Gamma\{\mu\}$ surrounds all admissible off-diagonal
solutions of the system \eq{TBE-z} and does not contain the points
$\pm\eta/2$ and $\{\lambda\}$. The existence of  such a contour follows from
Lemma \ref{main-lem}. The factor $1/N!$ appears due to
the invariance of the off-diagonal solutions with respect to the
permutations of $\{z\}$.

Let us observe finally that the integrand in \eq{Master-out} is
$i\pi$-periodic with respect to each $z_k$, and that it vanishes
as soon as any $z_k\to\pm\infty$. Hence, the sum of the residues of the
integrand within a set of strips of width $i\pi$ is equal
to zero. One can therefore evaluate the integral over the contour
$\Gamma\{\mu\}$ by taking the residues outside this contour
within this set of strips, i.e.
inside the contour $\Gamma\{\pm\frac{\eta}2\}\cup \Gamma\{\lambda\}$.
This leads to the master equation \eq{Master-2}.

          %%%%%%%%%%%%%    %%       %%     %%%%%%%%%%
          %%               %%%      %%     %%        %%
          %%               %%%%     %%     %%         %%
          %%%%%%%%         %%  %%   %%     %%         %%
          %%               %%   %%  %%     %%         %%
          %%               %%     %%%%     %%         %%
          %%               %%      %%%     %%        %%
          %%%%%%%%%%%%%    %%       %%     %%%%%%%%%%

\section{Thermodynamic limit}
\label{sec-th}

Starting from this section the parameters
$\lambda_1,\dots,\lambda_N$ correspond to the ground state
$|\psi(\{\lambda\})\rangle$ of the Hamiltonian in the subspace
${\cal H}^{(M/2-N)}$. This means in particular that
$\Im\lambda_j=0$ if $\Re\eta=0$, and $\Re\lambda_j=0$ if
$\Im\eta=0$. Recall also that it is enough to consider the case
$h\ge0$, what implies $N\le M/2$.

In \cite{KitMST04a} we showed how to obtain the thermodynamic limit of the
generating function ${\cal Q}_\kappa(m,0)$ from the master
equation by evaluating explicitly the integrals over $\Gamma\{\lambda\}$
and setting $d(z_k)=0$ in the remaining integrals. A similar method applies
to the time-dependent case, although the existence of the essential
singularities at $\pm\eta/2$ in the integrand  makes this procedure
more subtle.

Let us start with the representation \eq{Master-1}. Due to the symmetry of the
integrand  with respect to
the set $\{z\}$ we have
\be{Split-int}
\oint\limits_{\Gamma\{\pm\frac{\eta}2\}\cup \Gamma\{\lambda\}}
 \prod_{j=1}^{N}dz_j
=\sum_{n=0}^N C_N^n
\oint\limits_{\Gamma\{\pm\frac{\eta}2\}} \prod_{j=1}^{n}dz_j
\oint\limits_{\Gamma\{\lambda\}} \prod_{j=n+1}^{N}dz_j.
\ee
In this expression, the integral over the contour
$\Gamma\{\lambda\}$ can be rewritten as the
sum over partitions of the set $\{\lambda\}$ into two disjoint
subsets $\{\lambda\}=\{\lambda_{\alpha_+}\}\cup\{\lambda_{\alpha_-}\}$, with
$\#\{\lambda_{\alpha_+}\}=n$:
\begin{multline}\label{M-aster1}
 {\cal Q}_\kappa(m,t) =  \sum_{n=0}^N\frac{1}{n!}
\sum_{\{\lambda\}=\{\lambda_{\alpha_+}\}\cup\{\lambda_{\alpha_-}\}
\atop{|\alpha_+|=n}}
\oint\limits_{\Gamma\{\pm\frac{\eta}2\}}
 \prod_{j=1}^{n}\frac{dz_j}{2\pi i} \cdot
 \prod_{b=1}^{n}e^{itE(z_b)+imp(z_b)}
\prod_{b\in\alpha_-}^{n}e^{-itE(\lambda_b)-imp(\lambda_b)}
 \\
 \times
  \frac{\det_n \Omega_\kappa (\{z\} ,\{\lambda_{\alpha_+}\} |
\{z\}\cup\{\lambda_{\alpha_-}\})}
{ \prod_{a=1}^{n}{\cal Y}_\kappa (z_a| \{z\}\cup\{\lambda_{\alpha_-}\})}
 \cdot\frac{ \det_N \Omega (\{\lambda\} , \{z\}\cup\{\lambda_{\alpha_-}\}
  |\{\lambda\} )}
 { \det_N \Omega (\{\lambda\} , \{\lambda_{\alpha_+}\}\cup\{\lambda_{\alpha_-}\} | \{\lambda\})},
\end{multline}
where the elements in the sets $\{z\}\cup\{\lambda_{\alpha_-}\}$ and
$\{\lambda_{\alpha_+}\}\cup\{\lambda_{\alpha_-}\}$ are ordered accordingly.
Here we have used that
\begin{multline}\label{resH}
   \Res_{\{z_{n+1},\dots,z_N\}=\{\lambda_{\alpha_-}\}}
     \Bigl[ \det_N \Omega_\kappa ( \{z_1,\dots,z_N\} ,
 \{\lambda_{\alpha_+}\} \cup \{\lambda_{\alpha_-}\}|\{z_1,\dots,z_N\})\Bigr]\\
   = \prod_{a\in\alpha_-}
    {\cal Y}_\kappa (\lambda_a| \{z_1,\dots,z_n\} \cup \{\lambda_{\alpha_-}\} )
     \cdot
    \det_n \Omega_\kappa ( \{z_1,\dots,z_n\} , \{\lambda_{\alpha_+}\}
                           |\{z_1,\dots,z_n\} \cup \{\lambda_{\alpha_-}\}).
\end{multline}

Using now the system of Bethe equations \eq{BE_Y} for variables $\{\lambda\}$
we can present the integrand of \eq{M-aster1} as follows
\begin{multline}\label{transf-int}
\frac{\det_n \Omega_\kappa (\{z\} ,\{\lambda_{\alpha_+}\} |
\{z\}\cup\{\lambda_{\alpha_-}\})}
{ \prod_{a=1}^{n}{\cal Y}_\kappa (z_a| \{z\}\cup\{\lambda_{\alpha_-}\})}
\cdot\frac{ \det_N \Omega (\{\lambda\} , \{z\}\cup\{\lambda_{\alpha_-}\}
  |\{\lambda\} )}
 { \det_N \Omega (\{\lambda\} ,\{\lambda_{\alpha_+}\}\cup\{\lambda_{\alpha_-}\} | \{\lambda\})}
=\det_n\tilde M_{\kappa}(\{\lambda_{\alpha_+}\},\{z\})
\\
\times\frac{\prod\limits_{b=1}^n\prod\limits_{a\in\alpha_+}
\sinh(\lambda_a-z_b+\eta)\sinh(z_b-\lambda_a+\eta)}
{\prod\limits_{a,b\in\alpha_+}\sinh(\lambda_a-\lambda_b+\eta)
\prod\limits_{a,b=1}^n\sinh(z_a-z_b+\eta)}
\cdot
\frac{\det_N\Psi'(\{z\},\{\lambda_{\alpha_-}\}
|\{\lambda\})}
{\det_N\Phi'(\{\lambda_{\alpha_+}\}\cup\{\lambda_{\alpha_-}\})}.
\end{multline}
Here
\be{GFtiMjk}
(\tilde M_\kappa)_{jk}(\{\lambda_{\alpha_+}\},\{z\})
=t(z_k,\lambda_j)+\kappa t(\lambda_j,z_k)
\prod_{a\in\alpha_+}\frac{\sinh(\lambda_a-\lambda_j+\eta)}
{\sinh(\lambda_j-\lambda_a+\eta)}\cdot
\prod_{a=1}^n\frac{\sinh(\lambda_j-z_a+\eta)}
{\sinh(z_a-\lambda_j+\eta)},
\ee
and
\be{GFPhi}
\Phi'_{jk}(\{\lambda\})=\delta_{jk}\left[\left.\frac{d}{d\lambda}
\log\frac{d(\lambda)}{a(\lambda)}\right|_{\lambda=\lambda_j}
-\sum_{a=1}^{N} K(\lambda_j-\lambda_a)\right]+K(\lambda_j-\lambda_k),
\ee
with
\be{K}
K(\lambda)=\frac{\sinh2\eta}{\sinh(\lambda-\eta)\sinh(\lambda+\eta)}.
\ee
In $\Phi'(\{\lambda_{\alpha_+}\}\cup\{\lambda_{\alpha_-}\})$,
the columns are ordered such that the $n$ first ones correspond to the subset
$\{\lambda_{\alpha_+}\}$ and the $N-n$ last ones to $\{\lambda_{\alpha_-}\}$.
The matrix $\Psi'$ has a more complicated structure. For $k>n$ its entries
$\Psi'_{jk}$ coincide with the corresponding entries $\Phi'_{jk}$
in the $N-n$ last columns of
$\Phi'(\{\lambda_{\alpha_+}\}\cup\{\lambda_{\alpha_-}\})$. For the first
$n$ columns one has
\be{GCPsi}
\Psi'_{jk}= \frac{a(z_k)t(\lambda_j,z_k)-d(z_k)t(z_k,\lambda_j)
\prod\limits_{a=1}^{N}\frac{\sinh(z_k-\lambda_a+\eta)}
{\sinh(z_k-\lambda_a-\eta)}}
{a(z_k)+\kappa\, d(z_k)\prod\limits_{b=1}^n
\frac{\sinh(z_k-z_b+\eta)}{\sinh(z_k-z_b-\eta)}
\prod\limits_{b\in\alpha_-}
\frac{\sinh(z_k-\lambda_b+\eta)}{\sinh(z_k-\lambda_b-\eta)}},\qquad
k\le n
\ee

In the thermodynamic limit $M,N\to\infty$, $M/N=const$ this
expression can be simplified. Indeed, since $d(z)$ and $a(z)$
have zeros of order $M$ at $z=\eta/2$ and $z=-\eta/2$
respectively, one can show that the contributions of the corresponding terms to
the total result are bounded by $C^N/N!$. Therefore, at
$M,N\to\infty$, one can set $d(z)=0$ if $z$ is in the vicinity
of $\eta/2$ and $a(z)=0$  if $z$ is in the vicinity of
$-\eta/2$. This gives us a simplified representation for the
matrix elements \eq{GCPsi} for $k\le n$ in the thermodynamic
limit:
\be{GCPsinew}
\lim_{M\to\infty}\Psi'_{jk}=\widetilde \Psi'_{jk}=\left\{
\begin{array}{ll}
t(\lambda_j,z_k),&z_k\sim\frac\eta2,\non
-\kappa^{-1}t(z_k,\lambda_j)
\prod\limits_{a\in\alpha_+}\frac{\sinh(z_k-\lambda_a+\eta)}
{\sinh(z_k-\lambda_a-\eta)}\prod\limits_{b=1}^n
\frac{\sinh(z_k-z_b-\eta)}{\sinh(z_k-z_b+\eta)},
&z_k\sim-\frac\eta2.
\end{array}\right.
\ee
In fact we can say that the limiting value $\widetilde \Psi'_{jk}$ has
a cut between the points $\eta/2$ and $-\eta/2$.

The remaining steps are quite standard (see
\cite{KitMT00},\cite{KitMST02a}). In the thermodynamic limit the
distribution of the ground state parameters $\{\lambda\}$ can be
described by the spectral density $\rho_{tot}(\lambda)$. In its
turn the spectral density is a particular case of the
`inhomogeneous spectral density':
$\rho_{tot}(\lambda)=\left.\rho(\lambda,z)\right|_{z=\eta/2}$.
This inhomogeneous spectral density satisfies an integral
equation
\be{GFLiebeq}
 -2\pi i\rho(\lambda,z) +\int_{C}
K(\lambda-\mu)\rho(\mu,z)\,d\mu= t(\lambda,z),
\ee
where the integration contour $C$ in \eq{GFLiebeq} depends on the phase
of the model.  In the massless case $-1<\Delta\le1$ the contour $C$ is
an interval of the  real axis $[-\Lambda_h,\Lambda_h]$. The boundary
$\Lambda_h$ depends on the value of the magnetic field,
in particular $\Lambda_h\to\infty$ at $h\to0$.
For $\Delta>1$ ($\eta<0$) the
limits $\pm\Lambda_h$ are purely imaginary, more precisely the integral in
\eq{GFLiebeq} is taken over an interval of the imaginary axis. In
particular $\Lambda_h=-i\pi/2$ at $h=0$.

In the thermodynamic limit, one can compute the ratio of the determinants
$\det \widetilde \Psi'$ and $\det \Phi'$ in terms of the
inhomogeneous density. Indeed, since the last $N-n$ columns of the
matrix $\widetilde\Psi'$ coincide with the ones of the matrix
$\Phi'(\{\lambda_{\alpha_+}\}\cup\{\lambda_{\alpha_-}\})$,
we have
\be{ratio}
\frac{\det_N
 \widetilde\Psi'(\{z\},\{\lambda_{\alpha_-}\}|\{\lambda_{\alpha_+}\})}
{\det_N\Phi'(\{\lambda_{\alpha_+}\}\cup\{\lambda_{\alpha_-}\})}
=\det_n\left(\sum_{i=1}^{N}\left(\Phi'\right)^{-1}_{ji}
\widetilde\Psi'_{ik}\right).
\ee
Using the results of \cite{KitMT00} we have, for $z$ in a vicinity
of $\eta/2$,
\be{GClim+}
\sum_{i=1}^{N}\left(\Phi'\right)^{-1}_{ji}t(\lambda_i,z)\to
\frac{\rho(\lambda_j,z)}{M\rho_{tot}(\lambda_j)},\qquad
M\to\infty.
\ee
If $z$ is in a vicinity of $-\eta/2$,
we can set $z=\tilde z-\eta$ which gives $t(z,\lambda_j)=t(\lambda_j,\tilde z)$, hence
\be{GClim-}
\sum_{i=1}^{N}\left(\Phi'\right)^{-1}_{ji}t(z,\lambda_i)\to
\frac{\rho(\lambda_j,\tilde z)}{M\rho_{tot}(\lambda_j)}
=\frac{\rho(\lambda_j,z+\eta)}{M\rho_{tot}(\lambda_j)},\qquad
M\to\infty.
\ee
Thus, in the thermodynamic limit $M,N\to\infty$, $N/M=const$ one has
\be{ratio-TD-lim}
\frac{\det_N\widetilde\Psi'_{jk}(\{z\},\{\lambda_{\alpha_-}\}|
         \{\lambda_{\alpha_+}\})}
     {\det_N\Phi'_{jk}(\{\lambda_{\alpha_+}\}\cup\{\lambda_{\alpha_-}\})}
\to\det_n[{\cal R}^{\kappa}_n(\lambda_j,z_k|
\{\lambda_{\alpha_+}\},\{z\})],
\ee
where the function ${\cal R}^{\kappa}_n(\lambda,z|%
\{\lambda_1,\dots,\lambda_n\},\{z_1,\dots,z_n\})$
is defined differently in the vicinities of $\eta/2$ and $-\eta/2$:
\be{cal-R}
{\cal R}^{\kappa}_n(\lambda,z|
\{\lambda\},\{z\})=\left\{
\begin{array}{l}
\rho(\lambda,z),\qquad z\sim\eta/2;\\
-\kappa^{-1}\rho(\lambda,z+\eta)
\prod\limits_{b=1}^{n}\frac{\sinh(z-\lambda_b+\eta)
\sinh(z_b-z+\eta)}{\sinh(\lambda_b-z+\eta)\sinh(z-z_b+\eta)},
\qquad z\sim-\eta/2.
\end{array}\right.
\ee

It remains to replace in \eq{M-aster1} the sum over partitions
of $\{\lambda\}$ by integrals over the support of the spectral
density and we arrive at the multiple integral representation
for the dynamical correlation function of the third components
of spin:
\be{New-funct}
\langle\sigma_{1}^z(0)\sigma_{m+1}^z(t)\rangle=
2\langle\sigma_{1}^z(0)\rangle-1+2D^2_m
\left.\frac{\partial^2}{\partial\kappa^2}{\cal Q}_\kappa(m,t)\right|_{\kappa=1},
\ee
where
\begin{multline}\label{answer}
{\cal Q}_\kappa(m,t)=
\sum_{n=0}^{\infty}\frac{1}{(n!)^2}\int\limits_{-\Lambda_h}%
^{\Lambda_h} d^n\lambda
\oint\limits_{\Gamma\{\pm\frac\eta2\}}\prod_{j=1}^{n}
\frac{dz_j}{2\pi i}\cdot \prod_{a,b=1}^n\frac{
\sinh(\lambda_a-z_b+\eta)\sinh(z_b-\lambda_a+\eta)}
{\sinh(\lambda_a-\lambda_b+\eta)\sinh(z_a-z_b+\eta)}
          \\
\times
\prod_{b=1}^ne^{it(E(z_b)-E(\lambda_b))+im(p(z_b)-p(\lambda_b))}\;
\det_n\tilde M_{\kappa}(\{\lambda\},\{z\})
\cdot\det_n[{\cal R}^{\kappa}_n(\lambda_j,z_k|\{\lambda\},\{z\})].
\end{multline}
The contour $\Gamma\{\pm\frac\eta2\}$ surrounds the points
$\pm\frac\eta2$ and does not contain any other singularities of the integrand.
The parameter $\kappa$ in \eq{answer} is an arbitrary complex different from
$0,\infty$. The functions entering the integrand are defined in
\eq{E}, \eq{P}, \eq{GFtiMjk}, \eq{GFLiebeq} and \eq{cal-R}.

Due to the factors $\exp(itE(z_b))$ the integrand in \eq{answer}
has essential singularities in the points $\pm\frac\eta2$.
However, in the case $t=0$, these essential singularities
disappear and the integrals around $-\frac\eta2$ vanish. The
remaining part of the integrand has poles of  order $m$ at
$z_j=\frac\eta2$. Hence, at $t=0$ the sum over $n$ in
\eq{answer} is actually restricted to $n\le m$, and we reproduce
the result of \cite{KitMST02a} for the equal-time correlation
function of the third components of spin.

In Appendix \ref{free-f} we explain how one can deduce from this expression the
result obtained in \cite{ColIKT93} in the case of free fermions.

Using exactly the same method one can obtain an integral representation
for the $\sigma^z$ correlation function for the
partly inhomogeneous $XXZ$ chain, where we associate a set of
generic complex numbers $\xi_1,\dots,\xi_m$ with the first $m$ sites of the chain.
In this case one should replace in the result \eq{answer} the one-particle
momenta by their natural inhomogeneous modifications
\be{P-inh}
p_{inh}(\lambda)=\frac{i}{m}\sum_{k=1}^m
\log\left(\frac{\sinh(\lambda-\xi_k)}
{\sinh(\lambda-\xi_k+\eta)}\right).
\ee
The homogeneous case then corresponds to the limit $\xi_k=\eta/2$,
$k=1,\dots,m$.

          %%%%%%%%%%%%%    %%       %%     %%%%%%%%%%
          %%               %%%      %%     %%        %%
          %%               %%%%     %%     %%         %%
          %%%%%%%%         %%  %%   %%     %%         %%
          %%               %%   %%  %%     %%         %%
          %%               %%     %%%%     %%         %%
          %%               %%      %%%     %%        %%
          %%%%%%%%%%%%%    %%       %%     %%%%%%%%%%

\section{Conclusion}

In this article, we have obtained a multiple integral representation
for the dynamical $\sigma^z$ correlation function. It is clear, however,
that the method based on the master equation can be applied to other
dynamical correlation functions as well. In fact we have seen that the
time-dependent master equation for the generating function
${\cal Q}_\kappa(m,t)$ differs from its time-independent
analogue only by the factors $\exp(it(E(z)-E(\lambda))$, which automatically
appear in the framework of the form factor expansions for the correlation functions.
It is quite natural to expect that the same simple modification holds also
for other correlation functions.

One interesting further development would be to obtain a generalization of
the multiple integral
representations for the dynamical correlation functions at finite temperature.
A method to consider temperature correlation
functions by  algebraic Bethe ansatz was proposed recently in
\cite{GohKS04}. It is  possible that this  technique can
be successfully combined with the approach used in this paper.
In particular one obvious question is whether there exists also
a master equation for the temperature-dependent case. It would lead also to
the interesting question of the form factor expansion at non-zero temperature.

It is also well known that, for the case of free fermions $\Delta=0$, the
dynamical correlation functions of the $XXZ$ chain satisfy
difference-differential classical exactly solvable equations \cite{MccTW77,Per80,ItsIKS93}.
It is natural to wonder whether this property
holds also for general $\Delta$, or at least for some
specific cases. We hope that the multiple integral representations
for the dynamical correlation functions open a way to study this problem.

          %%%%%%%%%%%%%    %%       %%     %%%%%%%%%%
          %%               %%%      %%     %%        %%
          %%               %%%%     %%     %%         %%
          %%%%%%%%         %%  %%   %%     %%         %%
          %%               %%   %%  %%     %%         %%
          %%               %%     %%%%     %%         %%
          %%               %%      %%%     %%        %%
          %%%%%%%%%%%%%    %%       %%     %%%%%%%%%%

\section*{Acknowledgements}
J. M. M., N. S. and V. T. are supported by CNRS. N. K., J. M. M.,
V. T. are supported by the European network
EUCLID-HPRNC-CT-2002-00325.  J. M. M. and N.S. are supported
by INTAS-03-51-3350. N.S. is supported
by the French-Russian Exchange Program, the Program of RAS
Mathematical Methods of the Nonlinear Dynamics,
RFBR-02-01-00484, Scientific Schools 2052.2003.1.
N. K, N. S. and V. T. would like to thank the Theoretical Physics group
of the Laboratory of Physics at ENS Lyon for hospitality, which makes
this collaboration possible.

          %%%%%%%%%%%%%    %%       %%     %%%%%%%%%%
          %%               %%%      %%     %%        %%
          %%               %%%%     %%     %%         %%
          %%%%%%%%         %%  %%   %%     %%         %%
          %%               %%   %%  %%     %%         %%
          %%               %%     %%%%     %%         %%
          %%               %%      %%%     %%        %%
          %%%%%%%%%%%%%    %%       %%     %%%%%%%%%%

\appendix

\section{Admissible solutions of the twisted Bethe equations}%
\label{AS-TBE}

Let $e^{2z_j}=x_j$ and $e^\eta=q$. Then the system \eq{TBE-z}
takes the form
\be{TBE-P}
 Y_\kappa(x_j|\{x\})\equiv
(x_j-q^{-1})^M\prod_{a=1\atop{a\ne j}}^N(x_j-q^2x_a)-
\kappa q^{2N-2-M}\,(x_j-q)^M\prod_{a=1\atop{a\ne j}}^N(x_j-q^{-2}x_a)=0.
\ee
It is clear that, in the limit $\kappa\to 0$, all
admissible solutions of \eq{TBE-P} go to $q^{-1}$ and
the Jacobian matrix $(\partial Y_\kappa(x_j|\{x\})/
\partial x_k)$ has the rank zero at $\kappa=0$ and $x_j=q^{-1}$.
Our goal, however, is to solve the system \eq{TBE-P} for
$|\kappa|$ small enough, but not zero.

A simple example shows that, unlike in the inhomogeneous
case, the solutions $x_j(\kappa)$ of \eq{TBE-P} are not holomorphic
functions at $\kappa=0$. Indeed, let $q=i$ (free fermions). Since
for admissible solutions $x_j+x_k\ne 0$, we obtain
\be{Free-fer}
\left(\frac{x_j+i}{x_j-i}\right)^M=\kappa\,
e^{i\pi(N-1-M/2)}, \qquad\mbox{for}
\qquad q=i.
\ee
This system has an obvious solution
\be{Sol-free-fer}
x_j=i\frac{\theta_j+1}{\theta_j-1},\qquad\mbox{where}\qquad
\theta_j=-i|\kappa|^{\frac1M}e^{\frac{i\pi}{M}(2n_j+N-1)},\qquad
n_j\in\{0,1,\dots,M-1\}.
\ee
Thus, in this case, $x_j=x_j(\kappa^{\frac1M})$ and different choices
of the branch of $\kappa^{\frac1M}$ correspond to different solutions.

One can treat the general case similarly.
\begin{lemma}\label{adm-sol}
There exists $\kappa_0>0$ such that, for $0<|\kappa|<\kappa_0$,  all
admissible solutions of the system \eq{TBE-P} belong to the vicinity of
$q^{-1}$, but are separated from this point.
\end{lemma}

{\sl Proof.} Let us make in \eq{TBE-P} the substitution
$x_j=q^{-1}+\theta u_j$, where $\theta$ is one of the solutions of
$\theta^M=\kappa q^{2N-2-M}$. Then we arrive at
\begin{multline}\label{TBE-U}
\tilde Y_\theta(u_j|\{u\})\equiv
u_j^M\prod_{a=1\atop{a\ne j}}^N\Bigl(\theta (u_j-q^2u_a)+q-q^{-1}\Bigr)
\\
-(\theta u_j-q+q^{-1})^M\prod_{a=1\atop{a\ne j}}^N
\Bigl(\theta (u_j-q^{-2}u_a)+q^{-1}-q^{-3}\Bigr)=0.
\end{multline}
At $\theta=0$ one has
\be{theta-0}
u_j(0)=(q^{-1}-q)\cdot |q|^{\frac{2(1-N)}M}\cdot e^{\frac{2i\pi n_j}M},\qquad
n_j\in\{0,1,\dots,M-1\},
\ee
and
\be{Jac-0}
\left.\frac{\partial\tilde{Y}_\theta(u_j|\{u\})}{\partial u_k}
\right|_{u_j=u_j(0)\atop{\theta=0}}=M\delta_{jk}\cdot
u_j^{M-1}(0)\cdot(q-q^{-1})^{N-1}.
\ee
Hence, due to the implicit function theorem in a vicinity of
$\theta=0$, there exists  a unique holomorphic solution $u_j(\theta)$ of the
system \eq{TBE-U} which takes the value  \eq{theta-0} at $\theta=0$.
 This implies that
\be{behav-x}
x_j(\kappa)=q^{-1}+\theta u_j(0)+o(\theta).
\ee
Therefore, for $|\kappa|$ small enough, but not zero, all the
admissible solutions of the system \eq{TBE-P} are in a vicinity
of $q^{-1}$, but $x_j(\kappa)\ne q^{-1}$.
It is also evident that the
replacement $\theta\to\theta'=\theta e^{\frac{2\pi ik}M}$ corresponds simply to
a different choice of  the integers $n_j$ in \eq{theta-0}.$\Box$

Observe that, due to
\eq{behav-x}, the admissible  solutions
of the system \eq{TBE-P} for $|\kappa|$ small enough are off-diagonal
if the integers $n_j$ in \eq{theta-0} are pairwise distinct. Thus,
the system  \eq{TBE-P} (and therefore the system \eq{TBE-z}) has
$C^N_M=\dim({\cal H}^{(M/2-N)})$ different admissible off-diagonal
solutions. In the following two lemmas we prove that the corresponding
eigenstates $|\psi_\kappa(\{z\})\rangle$ are linearly independent.

\begin{lemma}\label{ortho}
Let the sets $\{z\}$ and $\{z'\}$ be two different
admissible off-diagonal solutions
of the system \eq{TBE-z} for $\kappa\ne0$. Then
\be{orth}
\langle \psi_\kappa(\{z'\})|\psi_\kappa(\{z\})\rangle=0.
\ee
\end{lemma}

{\sl Proof}. The scalar product \eq{orth} is given by \eq{scal-prod}.
If $z'_k\ne z_j, \quad \forall j,k$, then one can express
$\kappa\,d(z_k)$ in terms of $a(z_k)$ via \eq{TBE-z} for
the parameters $\{z\}$. Then we obtain
\be{vect-vect}
\langle \psi_\kappa(\{z'\})|\psi_\kappa(\{z\})\rangle=
\prod_{b=1}^{N}\Bigl(d(z'_b) a(z_b)\Bigr)
\frac{\prod\limits_{a,b=1}^N\sinh(z'_a-z_b+\eta)}
{\prod\limits_{a>b}^N\sinh(z_a-z_b)\sinh(z'_b-z'_a)}
\det_N\tilde M(\{z\},\{z'\}),
\ee
where $\tilde M(\{z\},\{z'\})=
\left.\tilde M_\kappa(\{z\},\{z'\})\right|_{\kappa=1}$ (see \eq{GFtiMjk}).
It was proved in \cite{KitMST02a} that
this matrix has an eigenvector with zero eigenvalue
\be{PSNev}
\sum_{k=1}^N\tilde M_{jk}v_k=0,\qquad\mbox{where}\qquad
v_k=\prod_{a=1}^n\sinh(z'_k-z_a)
\prod_{a=1\atop{a\ne k}}^n\sinh^{-1}(z'_k-z'_a).
\ee
Thus, in this case, the scalar product vanishes. If several
parameters $z'$ coincide with $z$, say, $z'_j=z_j$
for $j=1,\dots,n$, then one should first proceed to this limit in
the first $n$ columns of the matrix $\Omega_\kappa$ \eq{scal-prod},
and only in a second step use the equations \eq{TBE-z}. One can easily verify
that in this case the matrix we obtain has a zero eigenvector of the same type as
$v_k$. $\Box$

\begin{lemma}\label{n-ortho}
There exists $\kappa_0>0$ such that, for $0<|\kappa|<\kappa_0$,
\be{n-orth}
\langle \psi_\kappa(\{z\})|\psi_\kappa(\{z\})\rangle\ne0.
\ee
\end{lemma}
{\sl Proof}. The `square of the norm' of $|\psi_\kappa(\{z\})\rangle$
is proportional to the Jacobian \eq{norm}. The last one coincides with
the Jacobian \eq{Jac-0} up to a trivial factor which do not vanish at
$\kappa\ne0$.  Since $\partial\tilde{Y}_\theta(u_j|\{u\})/\partial u_k$
is a continuous function of $\theta$ and due to \eq{Jac-0}, it is
also non-vanishing
in a vicinity of $\kappa=0$.  $\Box$

Using these two lemmas we prove
\begin{thm}\label{basis}
There exists $\kappa_0>0$ such that, for $0<|\kappa|<\kappa_0$,  the
states $|\psi_\kappa(\{z\})\rangle$ corresponding
to the admissible off-diagonal solutions of the system
\eq{TBE-z} form a basis in the subspace ${\cal H}^{(M/2-N)}$.
\end{thm}
{\sl Proof}. It follows from Lemmas \ref{ortho}, \ref{n-ortho} that

1) Different admissible off-diagonal solutions of the system
\eq{TBE-z} correspond to different states  $|\psi_\kappa(\{z\})\rangle$.
Hence, the total number of the last ones is $\dim({\cal H}^{(M/2-N)})$.

2) These states are linearly independent. $\Box$

          %%%%%%%%%%%%%    %%       %%     %%%%%%%%%%
          %%               %%%      %%     %%        %%
          %%               %%%%     %%     %%         %%
          %%%%%%%%         %%  %%   %%     %%         %%
          %%               %%   %%  %%     %%         %%
          %%               %%     %%%%     %%         %%
          %%               %%      %%%     %%        %%
          %%%%%%%%%%%%%    %%       %%     %%%%%%%%%%

\section{Free fermions}\label{free-f}

In the particular case of free fermions $\Delta=0$ (i.e.
$\eta=-i\pi/2$), the  dynamical correlation function
$\langle\sigma_{1}^z(0)\sigma_{m+1}^z(t)\rangle$
can be written in the form \cite{ColIKT93}:
\be{CIKT}
\langle\sigma_{1}^z(0)\sigma_{m+1}^z(t)\rangle=
\left(\frac{2k_F}{\pi}-1\right)^2+
\frac{1}{\pi^2}\int\limits_{-k_F}^{k_F}
e^{4it\cos p+imp}\,dp\int\limits_{[-\pi,\pi]\setminus[-k_F,k_F]}
e^{-4it\cos q+imq}\,dq.
\ee
Here $\cos k_F=h/4$. This result was obtained by means of a summation
over the complete set of excited states. The constant term in \eq{CIKT}
corresponds to the diagonal contribution $\langle\sigma^z\rangle^2$.
The integration variable $p$ belongs to the
Fermi sphere $[-k_F,k_F]$ and describes the ground state distribution.
The integral over the parameter $q$ corresponds to the thermodynamic
limit of the sum over the excitations outside the Fermi sphere.

In this appendix, we show how to reproduce this result by the
direct use of the representation \eq{answer}.

At $\eta=-i\pi/2$ the integral equation \eq{GFLiebeq} is explicitly
solvable: the inhomogeneous spectral density has the form
\be{FF-rho}
\rho(\lambda,z)=\frac{i}{\pi\sinh2(\lambda-z)},
\ee
whereas the boundary of integration $\Lambda_h$
is defined by the relation $\cosh2\Lambda_h=4/h$. The
function ${\cal R}^{\kappa}_n$ at $\eta=-i\pi/2$ becomes
\be{FF-cal-R}
{\cal R}^{\kappa}_n(\lambda,z|\{\lambda\},\{z\})=\left\{
\begin{array}{l}
\frac{i}{\pi\sinh2(\lambda-z)},\qquad z\sim-i\pi/4;\\
\kappa^{-1}\frac{i}{\pi\sinh2(\lambda-z)}\qquad
z\sim i\pi/4.
\end{array}\right.
\ee
The main simplification, however, comes from the matrix $\tilde M_\kappa$, which
at $\Delta=0$ is proportional to $\kappa-1$:
\be{FF-tM}
(\tilde M_\kappa)_{jk}(\{\lambda\}|\{z\})=\frac{2(\kappa-1)}{\sinh2(\lambda_j-z_k)}.
\ee
Thus, after taking the second derivative over $\kappa$ and setting $\kappa=1$,
all the terms of the series \eq{answer} with $n>2$ vanish.

Let us first consider the term
${\cal Q}^{(2)}_\kappa(m,t)$ corresponding to
$n=2$. After differentiating with respect to $\kappa$ one has
\be{n=2}
\left.\frac{\partial^2}{\partial\kappa^2}{\cal Q}^{(2)}_\kappa(m,t)
\right|_{\kappa=1}=
\frac{1}{32\pi^4}\int\limits_{-\Lambda_h}^{\Lambda_h} \hspace{-2mm}d^2\lambda
\oint\limits_{\Gamma\{\mp\frac{i\pi}4\}}\hspace{-2mm}d^2z\cdot
\det\left(\frac{e^{itE(z_k)+imp(z_k)}}
{\sinh(\lambda_j-z_k)}\right)
\det\left(\frac{e^{-itE(\lambda_j)-imp(\lambda_j)}}
{\sinh(\lambda_j-z_k)}\right).
\ee
Like in the time-independent case the integral over $z_k$
can be taken by the residues outside of the contour
$\Gamma\{\mp\frac{i\pi}4\}$, i.e. in the points $z_k=\lambda_j$. This gives us
\be{n=2-cont1}
\left.\frac{\partial^2}{\partial\kappa^2}{\cal Q}^{(2)}_\kappa(m,t)
\right|_{\kappa=1}=
\frac{-1}{4\pi^2}\int\limits_{-\Lambda_h}^{\Lambda_h} d^2\lambda
\det\left(\frac{1-e^{it(E(\lambda_j)-E(\lambda_k))
+im(p(\lambda_j)-p(\lambda_k)}}
{\sinh(\lambda_j-\lambda_k)}\right).
\ee
It remains to take the second lattice derivative and, after the standard
change of variables $\cosh2\lambda_j=\cos^{-1}p_j$, we obtain
\be{n=2-cont2}
\left.2D^2_m\frac{\partial^2}{\partial\kappa^2}{\cal Q}^{(2)}_\kappa(m,t)
\right|_{\kappa=1}=\frac{4k_F^2}{\pi^2}-\frac1{\pi^2}
\left|\int_{-k_F}^{k_F}
e^{4it\cos p+imp}\,dp\right|^2.
\ee

The term ${\cal Q}^{(1)}_\kappa(m,t)$ corresponding to $n=1$ appears
to be more complicated. We have
\be{n=1}
{\cal Q}^{(1)}_\kappa(m,t)=\frac{\kappa-1}{4\pi^2}
\int\limits_{-\Lambda_h}^{\Lambda_h} d\lambda\left(
\oint\limits_{\Gamma\{-\frac{i\pi}4\}}+\kappa^{-1}\cdot
\oint\limits_{\Gamma\{\frac{i\pi}4\}}\right)
\frac{dz}{\sinh^2(\lambda-z)}\cdot
e^{it(E(z)-E(\lambda))+im(p(z)-p(\lambda))}.
\ee
Evaluating the integral over $\Gamma\{-\frac{i\pi}4\}$ we obtain
\begin{multline}\label{n=1-cont1}
\oint\limits_{\Gamma\{\frac{-i\pi}4\}}
\frac{dz}{\sinh^2(\lambda-z)}\cdot
e^{it(E(z)-E(\lambda))+im(p(z)-p(\lambda))}
\\
=-\oint\limits_{\Gamma\{\frac{i\pi}4\}}
\frac{dz}{\sinh^2(\lambda-z)}\cdot
e^{it(E(z)-E(\lambda))+im(p(z)-p(\lambda))}+2\pi[tE'(\lambda)+mp'(\lambda)].
\end{multline}
It is clear that the second lattice derivative of the last term
vanishes, and we have
\be{n=1-cont2}
D^2_m{\cal Q}^{(1)}_\kappa(m,t)=-\frac{(\kappa-1)^2}{4\pi^2\kappa}D^2_m
\int\limits_{-\Lambda_h}^{\Lambda_h} d\lambda
\oint\limits_{\Gamma\{\frac{i\pi}4\}}
\frac{dz}{\sinh^2(\lambda-z)}\cdot
e^{it(E(z)-E(\lambda))+im(p(z)-p(\lambda))}.
\ee
We can now explicitly differentiate this expression
with respect to $\kappa$ and $m$, which leads to
\be{n=1-cont3}
\left.2D^2_m\frac{\partial^2}{\partial\kappa^2}{\cal Q}^{(1)}_\kappa(m,t)
\right|_{\kappa=1}=\frac{4}{\pi^2}
\int\limits_{-\Lambda_h}^{\Lambda_h} d\lambda
\oint\limits_{\Gamma\{\frac{i\pi}4\}}
\frac{dz}{\cosh2\lambda\cosh2z}\cdot
e^{it(E(z)-E(\lambda))+im(p(z)-p(\lambda))}.
\ee
It remains to move the contour $\Gamma\{\frac{i\pi}4\}$ to the boundaries
of the strip
\be{move-cont}
\oint\limits_{\Gamma\{\frac{i\pi}4\}}dz=
\int\limits_{\mathbb{R}}\,dz-\int\limits_{\mathbb{R}+\frac{i\pi}2}\,dz,
\ee
and after the same change of variables as in the case $n=2$, we
finally
obtain
\be{n=1-cont4}
\left.2D^2_m\frac{\partial^2}{\partial\kappa^2}{\cal Q}^{(1)}_\kappa(m,t)
\right|_{\kappa=1}=\frac{1}{\pi^2}
\int_{-k_F}^{k_F}e^{4it\cos p+imp}\,dp\int_{-\pi}^{\pi}
e^{-4it\cos q+imq}\,dq.
\ee

Using now that\footnote{%
In fact one can obtain $\langle\sigma^z\rangle$ form the same generating
function ${\cal Q}_\kappa(m,t)$ by taking the first derivatives
with respect to $m$ and $\kappa$ at $\kappa=1$.}
$\langle\sigma^z\rangle=1-2k_F/\pi$ we reproduce the result \eq{CIKT}.

          %%%%%%%%%%%%%    %%       %%     %%%%%%%%%%
          %%               %%%      %%     %%        %%
          %%               %%%%     %%     %%         %%
          %%%%%%%%         %%  %%   %%     %%         %%
          %%               %%   %%  %%     %%         %%
          %%               %%     %%%%     %%         %%
          %%               %%      %%%     %%        %%
          %%%%%%%%%%%%%    %%       %%     %%%%%%%%%%

\end{document}